\renewcommand{\vec}[1]{\boldsymbol{#1}}

\documentclass[11pt,pra,floats,epsf,ifthen,showpacs,singlecolumn,superscriptaddress]{revtex4}%

\usepackage{graphicx}
\usepackage{amsmath}
\usepackage{color}

\definecolor{ngreen}{rgb}{0.2,0.6,0.2}

\definecolor{npurple}{rgb}{0.8,0.2,0.8}

\newcommand{\erf}[1]{Eq.~(\ref{#1})}

\begin{document}

\title{Qubit purification speed-up  for  three complementary continuous measurements}

\author{Rusko Ruskov}
\affiliation{Lundbeck Foundation Theoretical Center for Quantum
System Research, Department of Physics and Astronomy,   \AA{}rhus
University, DK-8000   \AA{}rhus   C, Denmark}
\affiliation{Laboratory for Physical Sciences,  8050 Greenmead Drive,   College Park, MD 20740, USA}

\author{Joshua Combes}
\affiliation{  Centre for Quantum Computation and Communication Technology (Australian Research Council),
Centre for Quantum Dynamics, Griffith University,
Brisbane, 4111, Australia}
\affiliation{Center for Quantum Information and Control, University of New Mexico, Albuquerque, NM 87131-0001, USA}

\author{Klaus M{\o}lmer}
\affiliation{Lundbeck Foundation Theoretical Center for Quantum
System Research, Department of Physics and Astronomy,   \AA{}rhus
University, DK-8000   \AA{}rhus   C, Denmark}
\author{Howard M. Wiseman}
\affiliation{  Centre for Quantum Computation and Communication Technology (Australian Research Council),
Centre for Quantum Dynamics, Griffith University,
Brisbane, 4111, Australia}
\date{\today}

\begin{abstract}
We consider qubit purification under simultaneous continuous measurement of
the three non-commuting qubit operators $\hat{\sigma}_x$, $\hat{\sigma}_y$, $\hat{\sigma}_z$.
The purification dynamics is quantified by   (i)   the average purification   rate,
and (ii)   the mean time of reaching given level of purity, $1-\varepsilon$.
Under ideal measurements (detector efficiency $\eta=1$), we show in the first case
an asymptotic   mean purification speed-up of $4$ as compared to a standard (classical)
single-detector measurement.   However by the second measure --- the mean time of first passage $\bar{T}(\varepsilon)$
of the purity --- the corresponding   speed-up is only $2$.
We explain these   speed-ups using
the isotropy of the qubit evolution that provides an equivalence between the original
measurement directions and three simultaneous measurements,   one with an axis  aligned along the Bloch vector
and the other   with axes  in the   two   complementary directions.
For inefficient detectors, $\eta=1-\delta <1$ the mean time of first passage $\bar{T}(\delta ,\varepsilon)$
increases since qubit purification competes with an isotropic qubit dephasing.
In the asymptotic high-purity limit ($\varepsilon, \delta \ll 1$) we show that
the increase possesses a {\it scaling behavior}: $\Delta \bar{T}(\delta,\varepsilon)$ is a function only
of the ratio  ${\delta}/{\varepsilon}$.
The increase $\Delta  \bar{T}({\delta}/{\varepsilon})$ is linear for small argument
but becomes exponential $\sim \exp({\delta}/2{\varepsilon})$
for   $\delta/\varepsilon$ large.
\end{abstract}

\pacs{quantum measurement, quantum control, quantum feedback, qubit, purification}
\maketitle

\section{Introduction}

Pure states are an important resource in quantum computation   and   communication  algorithms
 \cite{NielsenChuang,WisemanMilburn-control}.
While state purification is possible via {\it cooling}, this may be impractical for several reasons,
including long relaxation times, degenerate ground   states,   and   the   presence of several dephasing mechanisms.
A different purification process is possible via {\it continuous measurement}
when the (available to the observer) quantum state will purify continuously according to
the detector measurement result. Here, the speed at which one can  purify the state
is set by the detector measurement rate, and the final purity will depend on
how close  the detector is to a quantum-limited (100\% efficient) one.
This may become an important tool since
continuous measurements are also at the heart of various
quantum control applications\cite{WisemanMilburn-control},
including quantum  state stabilization via quantum feedback
 \cite{WisemanMilburn93,stable,RusKor-fb},
preparation of entangled states  \cite{RusKor-ent},
and for continuous error corrections \cite{errorz}.

In recent years several groups   have   suggested rapid purification protocols based on continuous
measurement and Hamiltonian feedback \cite{Jacobs,ComJac06,WisemanRalph,JordanKorotkov,ComWisJac08},
which   makes it possible    to considerably speed-up  purification.
For a single qubit, the problem was   first   analyzed by Jacobs \cite{Jacobs}, who recognized
that the   rate of average purification   can be enhanced using unitary transformation
at each measurement time step, so   as to make the state    always orthogonal   (in the Bloch sphere sense)
to the detector's  measurement  basis.
  This feedback algorithm (which has been
rigorously shown to be optimal  \cite{WisBou08}, { and rederived in Ref.~\cite{BelNegMol09}}) produces   a factor of 2
speed-up in the high-purity limit, when the state approaches the Bloch sphere surface.
  The speed-up is by comparison with a  no-feedback  measurement,
which can be completely understood classically  (see below).
This speed-up is a quantum mechanical effect since it is only possible if the system can exist
in superposition of the eigenstates of the measured observable.

Jacobs' protocol maximizes the average purification rate   in the sense that it
 minimizes the time $\tau$   it takes
the average purity $\langle p\rangle$   to reach a   certain level $1-\varepsilon$.
  A different, but equally well motivated, aim
was suggested by Wiseman and Ralph \cite{WisemanRalph}, that of minimizing
the mean time $\langle T\rangle$ for the purity $p$ to   attain a   certain level.
It was   argued   (and later rigorously confirmed \cite{WisBou08,BelNegMol09})
that the optimal protocol in this case
is to rotate the state so that at each measurement time
step it is aligned with the detector (i.e., making the state density matrix diagonal in the
basis of the detector observable).   This is the classical protocol
referred to above, and in the absence of any other dynamics, requires no feedback
to realize.

\begin{figure}
\vspace*{-0.3cm} \centering
\includegraphics[width=2.7in]{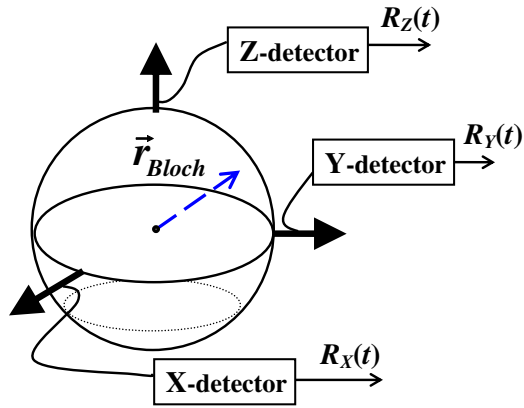}
\vspace*{-0.2cm}
\caption{A qubit measured by three orthogonal detectors.}
\label{schematic}
\end{figure}

In this paper we study a   recently suggested    purification protocol based on monitoring
of the qubit state via simultaneous continuous measurement of
the three non-commuting qubit operators \cite{RuskovKorotkovMolmer}
$\hat{\sigma}_x$, $\hat{\sigma}_y$, $\hat{\sigma}_z$  (Fig.\ref{schematic}).
(A related protocol
based on simultaneous measurement of these observables
was first considered in Ref.\cite{WeiNazarov}.)
%
%
In contrast to the above purification protocols there is no need to perform Hamiltonian feedback.
In this sense it is an ``open loop'' protocol,
{ similar to the ``random unitary control'' purification protocols introduced in Ref.~\cite{ComWisSco10}
for spin-$j$ systems.
Though operationally different, the two protocols can be shown to be equivalent in the special ($j=1/2$) case,
which is another consequence of the isotropy of the qubit evolution discussed below.}

Considering the rate of average purification   under ideal measurements (with detector efficiency $\eta=1$)
the purification speed-up is $4$,
 (this speed-up  is implicit   
in the results of Ref.~\cite{RuskovKorotkovMolmer}).
Here the  comparison is with a {\em single} measurement of the same strength with no feedback
(the speed-up of $4/3$ quoted in Ref.~\cite{ComWisSco10} comes from keeping the total measurement strength the same
in the two protocols). However,  we show here that
by the alternate metric ---   the mean time $\langle T\rangle$ of first passage   ---   the speed-up is   only   $2$.
  Both of these   purification speed-ups can be understood via the isotropy of the qubit evolution \cite{RuskovKorotkovMolmer}
in the Bloch space that allows   one   to represent the three detectors by another equivalent detector triad,
$\hat{\sigma}_{x'}$, $\hat{\sigma}_{y'}$, $\hat{\sigma}_{z'}$,
measuring at each time moment along the state and in the complementary directions.

We also study the purification dynamics for inefficient detectors, $\eta=1-\delta < 1$.
In the asymptotic high-purity limit ($\varepsilon, \delta \ll 1$)
 the increase $\Delta {  \bar{T}}(\delta ,\varepsilon)$ of the mean time of first passage $\langle T\rangle$
shows a {\it scaling behavior} (i.e. it depends only on the ratio   ${\delta}/{\varepsilon}$)
as seen from numerical calculations.
The increase   $\Delta {  \bar{T}}({\delta}/{\varepsilon})$   is linear for   relatively small inefficiency
but   grows exponentially, reaching   $\sim \exp({\delta}/2{\varepsilon})$
for   relatively   large inefficiency.

\section{Further background: Single detector purification protocols}

\subsection{Qubit impurity from single-detector measurement}

For a quantum-limited detector,
{ in the limit of infinite detector bandwidth, }
the evolution of the state $\rho(t)$ of a  quantum system
due to weak continuous measurements of a variable $X$ can be described
by the stochastic master equation (SME) \cite{WisemanMilburn-control}:
\begin{equation}
d\rho = \frac{\Gamma_0}{2}\, dt\, {\cal D}[X]\, \rho + \sqrt{\frac{\Gamma_0}{2}}\, dW(t)\, {\cal H}[X]\, \rho.
\label{SME}
\end{equation}
  Here   ${\cal D}[A]\rho \equiv A\rho A^{\dag} - \frac{1}{2}(A^{\dag} A \rho + \rho A^{\dag} A)$,
${\cal H}[A]\rho \equiv A \rho + \rho A^{\dag} - \mbox{Tr}[(A^{\dag} + A)\rho] \rho$,
and $dW$  is the increment of a Wiener noise process.
  The   internal Hamiltonian evolution of the system is either neglected
(in case of a large measurement strength $\Gamma_0$), or { eliminated}   
(see, e.g. Ref.~\cite{JordanKorotkov}).
The {\it measurement rate} $\Gamma_0$ sets    the   rate at which
information (about the final state) is extracted, and thus the rate at which the
system is projected onto a single eigenstate of $X$ \cite{Korotkov-99-01,VanStoMab0605}
(provided the spectrum of $X$ is non-degenerate).
We will specialize on the measurement of a qubit
and choose first $X=\hat{\sigma}_z$,   although   some results below are valid for a general $X$.

The measurement record in a small time interval $[t,t+dt)$ is given by
\begin{equation}
dR(t) =    \langle X \rangle\, dt + dW(t)/\sqrt{2 \Gamma_0}
\label{Result} ,
\end{equation}
where $\langle X \rangle =\mbox{Tr}[X \rho(t)]$,
and  $dW(t)$ is the same realization of the Wiener increment that appears in Eq.~(\ref{SME}),
thus  providing explicit dependence of the state evolution on the specific measurement result.
The evolution generated via Eq.~(\ref{SME}) corresponds to infinitesimal 
 measurement operators
(see, e.g. Refs. \cite{WisemanMilburn-control,JacobsSteck,JordanKorotkov}).
Since   the same observable $X$ is measured   at each time step, these   operators    
commute,   allowing one
 to integrate Eq.~(\ref{SME}) exactly.
Thus, for a finite time interval $[0,\tau)$ the final state will depend only on the
  time-averaged  measurement   record:
\begin{equation}
  \mu(\tau)   = \frac{1}{\tau} \int_0^{\tau} dt'  dR(t')
\label{R-integrated} .
\end{equation}
Introducing  the   
measurement operator   \cite{WisemanMilburn-control}
$$M_{\mu,z} \equiv \left( \frac{\Gamma_0 \tau}{\pi} \right)^{1/4}
\exp{[-(\mu - \hat{\sigma}_z)^2 \frac{\Gamma_0 \tau}{2} ]}$$
one obtains, using the method of un-normalized density matrices
(see, e.g. Refs. \cite{GoeGra94,Wis96,JacKni98,JacobsSteck}):
\begin{equation}
\rho(\tau,\mu) = \frac{M_{\mu,z}\rho(0)M_{\mu,z}^{\dag}}{  P(\mu) }
\label{POVM} .
\end{equation}
Here the probability distribution for ${  \mu}(\tau)$ is given by
  $P({  \mu})   = \mbox{Tr}[\rho M_{\mu,z}^{\dag} M_{\mu,z} ]$.

In the basis of the observable $X$, $\{|i\rangle \}$, the above result can be written as
a quantum Bayesian filter \cite{Caves,Korotkov-99-01}:
the update of the diagonal density matrix elements $\rho_{ii}$ will look exactly
as a classical Bayesian update of a ``probability distribution'':
$\rho_{ii}(\tau) = \rho_{ii}(0)   P(\mu|i)/P(\mu)  $, with Gaussian likelihoods $  P(\mu|i) $;
  this   was termed in Ref.~\cite{Korotkov-99-01}   the   {\it quantum-classical correspondence} principle.
The outcomes   $\mu$   will be Gaussian distributed around the eigenvalues of $X$, $\{ x_i \}$:
  $P (\mu|i) = \sqrt{\Gamma_0 \tau/\pi} \exp[-(\mu - x_i)^2 \Gamma_0\tau]$
with variance $\mbox{Var}= 1/2\Gamma_0\tau$,
and   expected   outcome probability
$
  P(\mu) = \sum_i \rho_{ii} P(\mu|i)
$.
The update of the non-diagonal matrix elements in (\ref{POVM}) will be according to the rule:
$\rho_{ij}(\tau) = \rho_{ij}(0) \sqrt{\rho_{ii}(\tau)\rho_{jj}(\tau) / \rho_{ii}(0)\rho_{jj}(0) }$
provided   the measurement is preformed   by a quantum-limited detector;
this implies that a pure state will remain pure \cite{WisemanMilburn-control}.  
From the above it is clear that the measurement time $\tau_{\rm meas}$  to distinguish
approximately   two eigenvalues, $x_i, x_j$ is set
by $\mbox{Var}|_{\tau_{  \rm meas}}=(x_i - x_j)^2/4$.
For a qubit this gives $  \tau_{  \rm meas}   =1/2\Gamma_0$ \cite{Korotkov-99-01}.

For further use we  also write down the quantum state evolution in terms of the
Bloch vector components of the state,
$x=2\mbox{Re}\rho_{12}$, $y=2\mbox{Im}\rho_{12}$, $z=\rho_{11}-\rho_{22}$.
From Eq.~(\ref{SME}) for the measurement of $\hat{\sigma}_z$ it follows:
\begin{eqnarray}
&& dz = (1-z^2)\, \sqrt{2\Gamma_0}\,\, dW
\nonumber \\
&& dx = - \frac{\Gamma_0}{\eta}\, x - z x\, \sqrt{2\Gamma_0}\,\, dW
\nonumber \\
&& dy = - \frac{\Gamma_0}{\eta}\, y - z y\, \sqrt{2\Gamma_0}\,\, dW
\label{Bloch-1-det}  .
\end{eqnarray}
In these equations we also included the effect of detector non-ideality (inefficiency) $\eta$,
 leading to a pure dephasing   :
In the ensemble averaged equations
(since   Eqs.(\ref{Bloch-1-det}) are in the   It\^{o}   form,   averaging means just to drop the noise term,
  and  corresponds to ignoring the detector results),
$\Gamma_0$ is the   decoherence rate due to an ideal detector of measurement rate $\Gamma_0$,
and $\eta =\Gamma_0/(\Gamma_0 + \gamma) \leq 1$ is the detector efficiency (ideality),
defined as the ratio of $\Gamma_0$ to the total decoherence $\Gamma = \Gamma_0 + \gamma$.
For a single detector the simplest model to describe the extra decoherence $\gamma$ is
to consider a second independent detector ``in parallel''
(i.e., measuring the same $\hat{\sigma}_z$ variable)
by adding terms similar to Eq.~(\ref{SME}) with $\Gamma_0$ replaced by $\gamma$,
and subsequently averaging over that detector output.
Thus, the density matrix available for an observer who takes into account only the results
of the first detector will be described by Eqs.(\ref{Bloch-1-det}).

\subsection{Single-detector purification protocols}

We consider first purification protocols via single-detector measurement with
an ideal (quantum-limited) detector.

\subsubsection{Purification without feedback}
For a single detector measurement without feedback, the state evolution in the detector basis
is essentially classical, Eq.~(\ref{POVM}).
That is, if the density matrix begins diagonal in the measurement basis (as it will be if it is a
completely mixed state), it remains so and
there is no way to distinguish $\{\rho_{11}(t),\rho_{22}(t)\}$ from a classical probability
distribution  \cite{Korotkov-99-01,RuskovKorotkovMizel}.
  For   continuous measurement one can consider the   purity or entropy   of the monitored state,
and in this particular case the von Neumann entropy of the state coincides with the Shannon entropy
(see, e.g. Refs. \cite{NielsenChuang,Tribus61}):
$S_{\rm  vN} = - \mbox{Tr}[ \hat{\rho}\,\ln \hat{\rho}] = - \left( \rho_{11} \ln{\rho_{11}} + \rho_{22} \ln{\rho_{22}} \right)$.
In what follows we will consider the so called linear entropy { (see, e.g. \cite{FuschJacobs01}),}  
$  s   = 1 - \mbox{Tr} \hat{\rho}^2 \equiv 1 - p$
which is   a monotonic function of $S_{\rm  vN}$.
Here  $p \equiv \mbox{Tr} \hat{\rho}^2 = \frac{(1 + x^2+y^2+z^2)}{2}$ is the  purity
expressed through the Bloch components of the state.

The corresponding equation for the purity { (at $\eta=1$)} follows from Eq.~(\ref{Bloch-1-det}):
\begin{equation}
dp = 2\Gamma_0\, \left[ (1-p) (1-z^2) \right] dt + 2 \sqrt{2\Gamma_0}\, z (1-p) \, dW
\label{Ito-dp-1det}  .
\end{equation}
Since purity $p$,   like entropy,   is invariant   under   unitary transformations,
without   loss of generality one can say that measurement   ``parallel to'' (in the same basis as)
the state
corresponds to $x=y=0$ and
\begin{equation}
dp = \Gamma_0\,[1-z^2]^2 dt + \sqrt{2\Gamma_0}\, z [1-z^2] \, dW
\label{Ito-dp0}  .
\end{equation}
so that the average change of purity is $\langle dp\rangle_{\parallel} = \Gamma_0\,(1-z^2)^2 dt$.
 We  note that the same result holds for non-ideal measurement, $\eta < 1$, since   the   detector's non-ideality
affects only the evolution of the non-diagonal density matrix elements;
 this means that
    a
non-ideal detector will purify the state if the measurement is along the  state.
It is clear that $\Gamma_0$ plays the role of a maximal classical purification (information acquisition) rate
that happen  when $z=\rho_{11}-\rho_{22}=0$, i.e. when the two outcomes are equally likely (see, e.g., Ref.~\cite{Tribus61}).
By approaching $z\rightarrow \pm 1$,   $\langle dp \rangle_{\parallel} \rightarrow 0$   since
``little information'' remains to be extracted
to clarify that collapse has happened  \cite{pre-existing}.
%
%

\subsubsection{Jacobs feedback purification protocol}

From   the discussion above,  it is intuitively   clear how the Jacobs' enhanced purification protocol works.
Given the state   $\rho(t)$,   one should continuously adapt the measurement basis
 [or, equivalently, rotate the state to  some   $\rho'(t)$]
so that the detector would perform measurement in a complementary direction with
respect to the eigenbasis of the rotated state (i.e., measuring {\it perpendicular to the state}, in the Bloch picture).
In the detector basis the rotated state again possesses {\it equally likely outcomes},   with $\rho'_{11}(t)=\rho'_{22}(t)$.
As   expected   classically \cite{Tribus61}, this procedure maximizes the average purification rate
since $p=p'$ under rotation, while  $z\rightarrow z'=0$  in Eq.~(\ref{Ito-dp-1det}).
  However,   the possibility to perform coherent rotations of the density matrix is of course a quantum mechanical effect.
The procedure also makes the purity evolution deterministic \cite{Jacobs}, i.e. the noise term
in Eq.~(\ref{Ito-dp-1det}) is zeroed so that
\begin{equation}
dp = 2\Gamma_0\,(1-p) dt
\label{dp_deterministic}
\end{equation}
and $\langle p\rangle = p$ under this protocol.
From this one can evaluate the time $\tau_{\perp}$ when the average purity $\langle p\rangle$ reaches
    a
given level $1-\varepsilon$:
\begin{equation}
\tau_{\perp} \simeq \frac{1}{2\Gamma_0} \int^{1-\varepsilon}\,\frac{dp}{1-p} \simeq \frac{1}{2\Gamma_0} \ln \varepsilon^{-1}
\label{tau_perp} .
\end{equation}
It should be noted that an attempt to evaluate the analogous time $\tau_{\parallel}$ in the case of the classical
measurement along the state, using $\langle dp\rangle_{\parallel}$  from  Eq.~(\ref{Ito-dp0}), will lead to a wrong
scaling of $\sim 1/\varepsilon$.
The reason is that
$\langle dp\rangle \neq d\langle{p}\rangle$ in general, and
the evolving purity distribution ${\cal P}(p,t)$ is generally different from
$\delta$-function  \cite{WisemanRalph}.
The correct evaluation of $\tau_{\parallel}$ is to calculate the average purity
$\langle p\rangle(t)$ by taking  into account the
exact solution, Eq.~(\ref{POVM}), of the stochastic evolution equations (\ref{SME}).
In the high-purity limit this leads to \cite{Jacobs,ComJac06,WisemanRalph,JordanKorotkov}
\begin{equation}
\tau_{\parallel} \simeq \frac{1}{\Gamma_0} \ln \varepsilon^{-1}
\label{tau_parallel}
\end{equation}
This is   exactly twice as long as the time in Eq.~(\ref{tau_perp}),
which establishes the speed-up of  2  for the Jacobs protocol.
 We note however, that, unlike the case of parallel measurement, the perpendicular measurement
will not purify to a completely pure state if the measurement is inefficient, as will be explored in Sec.~\ref{sec:3c}.
The reason is that the excess back-action in an inefficient perpendicular measurements results in
decay in the coherences of the state.

\subsubsection{Wiseman and Ralph purification protocol}

Instead of considering the time $\tau$   at which   the average purity $\langle p\rangle$ reaches a certain
level $1-\varepsilon$, one can also consider the average time $\langle T \rangle$
for a system to attain    that purity level,   $p(T)=1-\varepsilon$  \cite{WisemanRalph,WisBou08,ComWisJac08}.
It was noted in Ref.~\cite{WisemanRalph} that in many circumstances the
time $\langle T \rangle$ is
   a
more useful quantity.
The reason is that the time $T$ at which $p(T)=1-\varepsilon$, has a well-behaved
statistics \cite{WisemanRalph}, in contrast with $p(t)$ which has extremely long
tails at relatively small values of $p$.
  That is,   the averaged $\langle p\rangle$ is  strongly influenced by the rare cases
that are slow to purify.
Because of this there is a substantial disagreement between $\tau$ and $\langle T \rangle$
for a qubit.
It was shown in Ref.~\cite{WisemanRalph}, however, that good agreement is found between
$\langle T \rangle$ and $  T^{\rm log} {}$, defined as the time required for
 $\langle \ln[1-p(t)]\rangle$
%
to reach the certain level $\ln\varepsilon$.
This is because taking the logarithm de-emphasizes the tails, and indeed for a qubit
$\ln(1-p)$ has near-normal distribution  \cite{WisemanRalph}.

Therefore, we consider the stochastic equation for the   logarithm of the   linear
entropy  $s \equiv 1-p$.   It   follows from Eq.~(\ref{Ito-dp-1det})    that:
\begin{equation}
d\ln s = - 2\Gamma_0\, \{ 2 s + x^2 + 2 z^2 \} dt + 2 \sqrt{2\Gamma_0}\, z \, dW
\label{WR-dlog_s1}
\end{equation}
  (Here, without loss of generality we have   put $y=0$, since
      a
  single $\hat{\sigma}_z$-measurement
  keeps the state of the qubit in a fixed meridional plane).
In the Wiseman-Ralph feedback protocol \cite{WisemanRalph}
one keeps the monitored state along the detector $z$-axis
  (so $x^2=0$, while $z^2$ is maximized at $1-2s$).
Thus, the single detector purification is maximized
and in the high-purity limit, $s\approx 0$, we obtain
$\frac{\langle d\ln s \rangle_{\parallel}}{dt} \simeq - 4\Gamma_0$.
Using this,  the average time (of first passage) $\langle T \rangle$
is evaluated as $\langle T \rangle \simeq   T^{\rm log} {}$
and in the high-purity limit, $\varepsilon \ll 1$
\begin{equation}
    T^{\rm log}_{\parallel}   = \frac{1}{4\Gamma_0} \ln{\varepsilon^{-1}}
\label{time_pure1} ,
\end{equation}
which is   half the size of $\tau_{\perp}$,   and
4 times shorter than $\tau_{\parallel}$. %
  In   what follows we will use these results to understand the purification
via three simultaneous complementary measurements without feedback.

\section{Continuous measurement of three complementary qubit variables}

We consider continuous measurement of the qubit complementary observables
$\hat{\sigma}_x$, $\hat{\sigma}_y$, and $\hat{\sigma}_z$  by three
independent linear detectors (see Fig. \ref{schematic}).
In principle, given a $\hat{\sigma}_z$-detector, the measurement
of, say, $\hat{\sigma}_x$  can be implemented operationally
by performing   
fast unitary rotation, $U_{x}(t)$, of the state towards the $x$-axis
``at the beginning'' of the measurement interval $dt$,
then a continuous measurement with the $z$-detector
and finally, a  backwards transformation, $U_{x}^{-1}(t+dt)$ ``at the end'' , in any
infinitesimal measurement time step $dt$,   as discussed in Ref.~\cite{ComWisSco10}.
We also mention that a simultaneous measurement of the three observables
is possible to implement, at least in principle,
in a quantum optics setup \cite{RuskovMolmer-unpublished}
(e.g., using a single atom in an optical cavity field).

In the case of detectors measuring the qubit in the  bases of
non-commuting observables, it is
impossible to use the quantum Bayes rule (\ref{POVM})
for finite times because
the measurement back-actions, $M_{\mu,x}$, $M_{\mu,y}$, $M_{\mu,z}$,
do not commute with each other.
The measurement back-actions commute only for small
measurement time intervals $dt$ so that one should apply
the POVM update in its differential form
Eq.~(\ref{SME}) and then sum up the
contributions to the qubit evolution.
For the case of mutually unbiased observables
$\hat{\sigma}_x$, $\hat{\sigma}_y$, and $\hat{\sigma}_z$
it is convenient to use the Bloch vector components,
$\vec{r}=(x,y,z) = \{\mbox{Tr}[\hat{\rho}\hat{\sigma}_k], k = x,y,z\}$ determined by the
  time-dependent   qubit density matrix $\hat{\rho}$.
From Eq.~(\ref{Bloch-1-det}) for the influence due to $\hat{\sigma}_z$-measurement,
the influences due to $\hat{\sigma}_x$ ($\hat{\sigma}_y$)-measurement can be
obtained  by cyclic replacements $z \rightarrow x \rightarrow y$,
bearing in mind that the variables
$x$($y$) should play the same role as $z$
in a $\hat{\sigma}_z$-measurement.
In this way we obtain
the following evolution equations in   It\^{o}   form (see Ref.~\cite{RuskovKorotkovMolmer})
that take into account the three measurement records $dR_k(t), k=x,y,z$ as in Eq.~(\ref{Result}),
and correspondingly introduce three independent Wiener processes, $dW_i\, dW_k = \delta_{ik}\, dt$:
\begin{eqnarray}
&& dx = - (\Gamma_y + \Gamma_z)\,x + (1-x^2)\, \sqrt{2\Gamma_{0,x}}\,\, dW_x
\nonumber\\
&&\quad\qquad  - x\,y\, \sqrt{2\Gamma_{0,y}}\,\, dW_y - x\,z\, \sqrt{2\Gamma_{0,z}}\,\, dW_z
\label{3det-Itox}\\
&& dy = - (\Gamma_z + \Gamma_x)\,y + (1-y^2)\, \sqrt{2\Gamma_{0,y}}\,\, dW_y
\nonumber\\
&&\quad\qquad  - y\,z\, \sqrt{2\Gamma_{0,z}}\,\, dW_z - x\,y\, \sqrt{2\Gamma_{0,x}}\,\, dW_x
\label{3det-Itoy}\\
&& dz = - (\Gamma_x + \Gamma_y)\,z + (1-z^2)\, \sqrt{2\Gamma_{0,z}}\,\, dW_z
\nonumber\\
&&\quad\qquad  - x\,z\, \sqrt{2\Gamma_{0,x}}\,\, dW_x - y\,z\, \sqrt{2\Gamma_{0,y}}\,\, dW_y
\label{3det-Itoz}
\end{eqnarray}
Here $\Gamma_k \equiv \gamma_k + \Gamma_{0,k}$ are the total decoherence rates for each detector,
including individual dephasings $\gamma_k$ and measurement rates $\Gamma_{0,k}, k=x,y,z$.
  Note   that a pure dephasing,   say in the $x$-basis,
 causes the decay of the $z$- and $y$-components.
Eqs.~(\ref{3det-Itoy}) and (\ref{3det-Itoz}) are  simply obtained by
cyclic permutation of the variables in Eq.~(\ref{3det-Itox}).

\subsection{Qubit evolution with identical detectors}

\begin{figure}
\vspace*{-0.3cm} \centering
\includegraphics[width=2.7in]{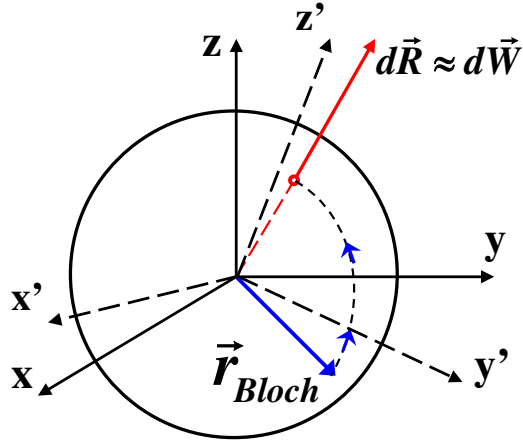}
\vspace*{-0.2cm}
\caption{Bloch equations invariance under rotations for three complementary measurements with identical detectors.
The evolution is determined by the relative orientation of the Bloch vector and
the vector of results \cite{RuskovKorotkovMolmer}.}
\label{Fig2}
\end{figure}

In what follows we consider the simplest (but still   rich)   case of
three identical detectors: $\Gamma_{0,k} = \Gamma_{0}$ and
$\gamma_k = \gamma\geq 0$.
Then the qubit evolution
(\ref{3det-Itox})--(\ref{3det-Itoz}) can be rewritten in a vector form as
\begin{eqnarray}
&& d {\vec{r}} = -2 \Gamma\,\vec{r}\,   dt
+   \sqrt{2\Gamma_{0}}\,\, \left\{\vec{dW}(t)\,(1-r^2)
- \left[\vec{r}\times
\left[\vec{r}\times \vec{dW}(t)\right]\right] \right\}
\label{Ito-vec2} ,
\end{eqnarray}
where $\vec{dW}(t) \equiv \{dW_x,dW_y,dW_z\}$ is the vector of Wiener increments
corresponding to the vector of results $\vec{dR}=\{dR_x,dR_y,dR_z\}$, similar to Eq.~(\ref{Result}).
The evolution (\ref{Ito-vec2}) is invariant under
arbitrary rotations  (see Fig. \ref{Fig2})   of the coordinate system in Bloch space \cite{3D-Gaussian}
%
(as   is   the ensemble averaged evolution: $\dot{\vec{r}}=-2\Gamma\, \vec{r}$).
  Hence    while measurement of only
    a
  single observable
$\hat{\sigma}_k$ ``attracts'' the qubit state to one of the
corresponding eigenvectors, the simultaneous measurement of
$\hat{\sigma}_x$, $\hat{\sigma}_y$, and $\hat{\sigma}_z$ leads to no
preferable direction in the Bloch space.
In Ref.~\cite{RuskovKorotkovMolmer} this isotropy of the qubit evolution
was shown to lead to a locally isotropic Brownian diffusion of the
direction of the Bloch vector.
In particular, for pure states under ideal measurements, the point
on the Bloch sphere diffuses isotropically  \cite{Perrin} 
with a diffusion coefficient $\Gamma_0$  \cite{RuskovKorotkovMolmer}.
The evolution isotropy was used then to construct simple quantum state estimations \cite{RuskovKorotkovMolmer}.
%
In what follows, we will use the qubit evolution isotropy to understand the
qubit purification dynamics.

\subsection{Purification dynamics with three complementary measurements}

Eq.~(\ref{Ito-vec2}) shows that the measurement term component  along the radius
is $\sqrt{2\Gamma_{0}}\, (\vec{dW}\cdot\vec{e_r}) \, (1-r^2)$ and vanishes on the sphere.
The decreasing $r$-dependent coefficient
suggests that $r=1$ is an attractor of the random evolution.
Thus, the qubit state will purify in the ideal case ($\gamma=0$)
for any realization of the measurement process.

To establish further the purification dynamics, we start with
Eq.~(\ref{Ito-vec2}) and transform it to polar coordinates, $r$, $\theta$, $\phi$.
The equation for the  radius decouples from the other two
and reads \cite{RuskovKorotkovMolmer}:
\begin{equation}
dr = 2\Gamma_0\, (1/r-r/\eta )dt  + \sqrt{2\Gamma_0} (1-r^2) \, dW_r
\label{Ito-radial} ,
\end{equation}
where $dW_r \equiv \vec{dW}\cdot\vec{e_r} $ is the projection of the noise vector
onto the Bloch vector direction,
$\vec{e_r}=(\sin\theta \cos\varphi,\sin\theta \sin\varphi,\cos\theta)$,
and the normalization of the noise is $dW_r dW_r =dt$.

As seen from Eq.\ (\ref{Ito-radial}), if $r< \eta^{1/2}$, then on
average $\dot{r}>0$, which means state purification for ideal measurements, $\eta=1$.
Since   the   equation for $r$ is singular at the origin, it will be more convenient
to consider further the dynamics of the   purity   $p=\mbox{Tr}\hat{\rho}^2 = \frac{(1+r^2)}{2}$.
 In   this choice $1/2 < p < 1$, i.e., a totally mixed state corresponds to $p=1/2$.
We   find
\begin{eqnarray}
&&dp = 2\Gamma_0\, \{   1-(2 p - 1)/{\eta}   + 2 (1-p)^2  \}   dt
+   2 \sqrt{2\Gamma_0} (1-p) \sqrt{2 p -1}\, dW_r
\label{Ito-dp-3det}  .
\end{eqnarray}
For an ideal measurement, $\eta=1$,
and starting from a non-pure initial state, the qubit will purify
($  p  \rightarrow 1$) on a time scale of the order of
$\tau_{\rm meas} \sim \Gamma_0^{-1}$. For a non-ideal measurement, $\eta < 1$, purity will
continue to fluctuate around a stationary average value,
$\langle p\rangle_{\rm   st} < 1$.

\subsubsection{Fokker-Planck equation for purity}

To quantify the purity dynamics and asymptotic purity distribution
we first consider the Fokker-Planck equation (FPE) corresponding to
Eq.~(\ref{Ito-dp-3det}).
Using the standard coefficients (see Ref.~\cite{Gardiner}):
\begin{eqnarray}
&& A(p) = 2\Gamma_0\, [    1-(2 p - 1)/{\eta}   + 2 (1-p)^2 ]
\label{FP-A-coeff}\\
&& B(p) = 8\Gamma_0\, ( 2 p - 1 ) (1-p)^2
\label{FP-B-coeff}
\end{eqnarray}
the FPE reads
$$\frac{\partial {\cal P}(p,t)}{\partial t} =
-\frac{\partial}{\partial p}
\left[ A(p) {\cal P}(p,t)\right] +
\frac{1}{2}\,\frac{\partial^2}{\partial p^2}\,
\left[ B(p) {\cal P}(p,t)\right],$$
and   the initial distribution is taken to be
${\cal P}(p,0) = \delta(p-p_0)$.
   At $t \gg \tau_{\rm meas}$ the purity  reaches a
stationary distribution \cite{RuskovKorotkovMolmer}
\begin{equation}
{\cal P}_{\rm   st}(p,\eta) =
N^{-1}\, \frac{\sqrt{2p-1}}{(1-p)^3}\,
\exp{\left[-\frac{(2p-1) (1-\eta)}{2(1-p)\eta}\right]} ,
  \label{FP-Pur-stationary}
\end{equation}
where $N$ is the normalization.
    For $\eta\rightarrow 1$, the stationary distribution ${\cal P}_{\rm   st}(p,\eta)$ approaches the
$\delta$-function at $p=1$.

The purification dynamics can be approximated using the
ensemble-averaged purification rate  \cite{Jacobs,WisemanRalph}
obtained from the above It\^{o}   equation    (\ref{Ito-dp-3det})
 for   purity:
\begin{equation}
\langle dp \rangle = 2\Gamma_0\,   \langle 1-(2 p - 1)/{\eta} + 2 (1-p)^2  \rangle   \, dt
\label{av-pur-rate}  .
\end{equation}
Now consider a naive { approach} to integrating  Eq.~(\ref{av-pur-rate}), in which we
replace $p$ by $\langle p\rangle$ everywhere it appears on the right-hand-side:
\begin{equation}
\langle dp \rangle = 2\Gamma_0\,   \left[ 1-(2 \langle p \rangle - 1)/{\eta} + 2 (1-\langle p\rangle )^2  \right] \, dt
\label{av-pur-rate-naive}  .
\end{equation}
Contrary to the single-detector measurement, it can be shown,
by numerically solving the FPE, that the evolution of the average purity
$\langle p\rangle$  obtained  from the naive Eq.~(\ref{av-pur-rate-naive})   is very close to the true average
$\langle p(t)\rangle_{\rm   FP}  = \int_{1/2}^1 p\, {\cal P}(p,t) \, dp$,
obtained from FPE.   (The reasons for this difference will be discussed below.)
Numerically, the approximation is    best  in the high purity limit ($\langle p \rangle=1 - \langle s\rangle$,
$\langle s\rangle\ll 1$) and
for almost ideal detectors, $\eta \simeq 1$. In the high-purity limit we can discard terms of order $\langle s\rangle^2$
(i.e. the final term)
so that  \erf{av-pur-rate-naive} becomes a simple linear equation. For   ideal measurements the time when the average
purity reaches the level of $1-\varepsilon, \varepsilon \ll 1$ is   thus
\begin{equation}
\tau_{\rm   iso} \simeq \frac{1}{4\Gamma_0} \int^{1-\varepsilon} dp \frac{1}{1-p} = \frac{1}{4\Gamma_0} \ln{\varepsilon^{-1}}
\label{3det-tau-pur} ,
\end{equation}
which is four times shorter than the standard time $\tau_{\parallel}$ of classical measurements, Eq.~(\ref{tau_parallel}).
This means   with three orthogonal measurements we obtain    4 times speed-up as compared to 2 times speed-up of
the Jacobs   (single measurement)   purification protocol.

This result can be understood as follows. By the isotropy of the qubit evolution under three complementary
measurements, in each time moment $t$ one can represent the measurements in the $x$, $y$, $z$ directions
by measurements in an equivalent triple of directions  $x'$, $y'$, $z'$
(in the Bloch space, Fig. \ref{Fig2}),   chosen so that
$z'$ is   parallel to   the state (i.e., in the basis of $\hat{\sigma}_{z'}$ the density matrix $\rho(t)$ is diagonal),
while directions $x'$, $y'$ are perpendicular to the state;   of course these chosen directions must change
in time   according
to the evolution of the state, $\rho(t)$.
The measurements in the $x'$, $y'$ directions in the time interval $[t,t+dt)$
are termed ``good'' measurements as they are precisely the unbiassed measurements of Jacobs that maximize
the average purification rate, and the $z'$-measurement along the state would be termed ``bad'' by obvious reason.
This observation could be confirmed by rewriting Eq.~(\ref{av-pur-rate}) for $\eta=1$ in the following way:
\begin{equation}
\langle dp \rangle = 2\Gamma_0\,   \langle    (1-p) + (1-p) + 2 (1-p)^2    \rangle   dt
\label{av-pur-rate1} ,
\end{equation}
where the first two terms correspond to the deterministic change, $dp$, Eq.~(\ref{dp_deterministic}),
and the last term coincides with the average change of $\langle dp \rangle_{\parallel}$ due to
measurement along the state, Eq.~(\ref{Ito-dp0}).
In the high purity limit, the last term is suppressed and the two ``good'' measurements
contribute each a speed-up of 2 that amount to a total speed-up \cite{note1} of 4.
  Moreover, this makes clear why the naive approach (replacing $\langle p^2 \rangle$ by $\langle p\rangle^2$)
to calculating the mean purity works
for the isotropic measurement: Jacob's protocol gives deterministic growth of the purity, so that
$\langle p^2 \rangle = \langle p\rangle^2$. For isotropic measurement the purification in the mean is dominated
by the perpendicular measurements, as just shown. Therefore the purification is approximately deterministic,
and becomes more so the purer the state becomes. By contrast, with a single measurement and no control
(the parallel measurement case), the purification is far from deterministic. In particular the long tail of low purities
means it is impossible to
obtain accurate results by replacing $\langle p^2 \rangle$ by $\langle p\rangle^2$ \cite{WisemanRalph}.


The ``splitting'' of the measurements into ``good'' and ``bad'' is also applicable when
the goal of minimizing the mean time $\langle T\rangle$ is considered.
As in the single-detector case~\cite{WisemanRalph}, we take the log-entropy evolution, $\ln{s},\, s=1-p$;
similarly to that case $\ln s$ should have more symmetric distribution and therefore
its average rate of change would correspond to the mean time of first passage (MTFP).
Using   It\^{o}   equation for three ideal detections, the average change
of log-entropy in the high-purity limit ($s\to 0$) reads:
\begin{equation}
\langle d\ln s\rangle \simeq - 8\Gamma_0\, dt = -(4+2+2)\Gamma_0\, dt
\label{Ito-dlog_s3}  ,
\end{equation}
i.e., the three detector measurements now ``splits'' into
one ``good'' measurement ($z'$), directed along the state, which gives the rate $4\Gamma_0$
as in Eq.~(\ref{time_pure1}) [Wiseman and Ralph protocol],
and two ``bad'' measurements ($x'$, $y'$), directed perpendicular to the state, that give one-half
of this rate each, see Eq.~(\ref{tau_perp}).
Correspondingly, the mean time $\langle T\rangle \simeq   T^{\rm log}$
for the average log-entropy $\langle \ln s\rangle$ to reach $\ln\varepsilon$ is given by
\begin{equation}
  T^{\rm log}_{\rm   iso} = \frac{1}{8\Gamma_0} \ln{\varepsilon^{-1}}
\label{mean-time-3det}  .
\end{equation}
This is two times shorter than the mean time for a single-detector no-control
measurement, $  T^{\rm log}_{\parallel}$, Eq.~(\ref{time_pure1}).
Therefore,   in comparison   to the   parallel single measurement case,
  three complementary   measurements give a speed-up of 2   in terms of the
mean time to attain a given purity. We will verify this result by explicitly
calculating the MTFP,  in the following subsection where we consider non-ideal detectors.

\subsection{Purification dynamics for non-ideal detectors} \label{sec:3c}

It is now interesting to ask the question:
Given   non-ideal detectors with   $\eta \equiv 1- \delta$,
what level of purity, $p = 1-\varepsilon$
can be reached and what time is needed?
We now consider in detail this question, with emphasis  on the high-purity   and high-efficiency
limit, when $\varepsilon \ll 1$ and $\delta \ll 1$.
The answer to the above question will depend on the goal
examined under purification.

\subsubsection{The goal of   having the average purity   reach    the level $(1-\varepsilon)$}

Since the average purity $\langle p(t)\rangle_{  \rm FP}$ described by the FPE
is numerically close to that from the   naive ensemble-average evolution
Eq.~(\ref{av-pur-rate-naive}),
$\langle p(t)\rangle$
  (see also Ref.\cite{RuskovKorotkovMolmer}),
we will use the    latter
in our analysis.
In particular, the stationary value $\langle p\rangle_{\rm   st}$ is close to the true stationary
value, obtained from the stationary distribution, Eq.~(\ref{FP-Pur-stationary}), and reads:
\begin{equation}
\langle p\rangle_{\rm   st} = 1 + \frac{1}{2} \left( \frac{1}{\eta} - \sqrt{\frac{1}{\eta^2} + \frac{2}{\eta} -2} \right)
\label{p-stationary}  .
\end{equation}
In the high-ideality limit it gives $\langle p\rangle_{\rm   st} \simeq 1 - \frac{\delta}{2}$.
Therefore, $\langle p\rangle$ can reach a purity level $(1-\varepsilon)$ only if
\begin{equation}
\varepsilon \geq \frac{\delta}{2}
\label{p-bound}  .
\end{equation}
The time for $\langle p\rangle$ to reach $1-\varepsilon$ can be calculated from Eq.~(\ref{av-pur-rate-naive}),
and in the high-purity limit it gives:
\begin{equation}
\tau_{\rm   iso} { (\delta,\varepsilon)} \simeq \frac{1}{4 \Gamma_0} \left[ \ln  \frac{1}{2\varepsilon}   - \ln\left(1-\frac{\delta}{2\varepsilon}\right)\right]
\label{p-tau-pure-delta}  .
\end{equation}
Here we note that the purification rate can be again understood   by   splitting into ``good'' and ``bad'' measurements.
Indeed, the single detector average purification for $\eta < 1$ can be calculated from Eq.~(\ref{Bloch-1-det}):
\begin{equation}
\langle dp \rangle = 2\Gamma_0\,   \left\langle (1-p) (1-z^2) + \frac{1}{2}\left(1-\frac{1}{\eta}\right) (2p-1-z^2)\right\rangle   dt
\label{Itoav-dp-1det-eta}  ,
\end{equation}
and one can observe that $\langle dp \rangle$ is  maximized for a measurement in perpendicular direction
(${z}^2=0$)
for not too small $\eta$ (e.g.   even   for a totally mixed state, $p=1/2$, this happens for $\eta > 1/2$).
Thus, the first term in Eq.~(\ref{av-pur-rate}) comes exactly from two ``good'' measurements
(perpendicular to the state), $2\langle dp \rangle_{\perp} = 1- (2p-1)/\eta$,
while the second term comes from the ``bad'' measurement (parallel  to the state) as   in Eq.~(\ref{Ito-dp0}),
 and is suppressed in the high-purity limit. 
%
%
%
Therefore, the asymptotic result for $\tau_{\rm   iso}{ (\delta,\varepsilon)}$ comes entirely from the two good measurements,
as in the ideal case.
The time increase
in Eq.~(\ref{p-tau-pure-delta})
with respect to the ideal measurement, Eq.~(\ref{3det-tau-pur}) is linear
in the ratio $\delta/\varepsilon$ for  $\delta \ll \varepsilon$ and
  diverges { logarithmically} to $\infty$ as the impurity level $\varepsilon$ approaches the bound, Eq.~(\ref{p-bound}).
To reach impurity level $\varepsilon < \frac{\delta}{2}$ is simply impossible.

\subsubsection{The goal of   having a certain mean time $\langle T\rangle$ at which the  purity $p(T)$ reaches    the level $( 1-\varepsilon )$}

We now proceed with the  investigation of the mean time $\langle T\rangle$ for the purity to reach $p(T)=1-\varepsilon$
for three complementary non-ideal measurements.
Following Ref.~\cite{WisemanRalph} we write
the   It\^{o}   equation for the log-entropy:
\begin{eqnarray}
&& d\ln s\mid_{\rm   iso} = - 2\Gamma_0\, \left\{ 2 - 2 s + \frac{2}{\eta} + \left( 1-\frac{1}{\eta}\right)\frac{1}{s} \right\} dt
  +  2 \sqrt{2\Gamma_0} \sqrt{1 - 2 s}\, dW_r
\label{Ito-dlog_s3a} .
\end{eqnarray}
Similar to the ideal case, the average change of log-entropy  can be represented as a sum of
the measurement along the state and two measurements  perpendicular to the state.
Indeed, the change $d\ln s$ for a single non-ideal detector can be written as:
\begin{eqnarray}
&& d\ln s = - 2\Gamma_0\, \left\{ 2 s + x^2 + 2 z^2 + \left( 1-\frac{1}{\eta}\right)\frac{x^2}{2s}\right\} dt
  +   2 \sqrt{2\Gamma_0} z \, dW
\label{dlog_s1_eta}  .
\end{eqnarray}
We note that for given entropy $s = 1 - p$, the average change $\langle d\ln s \rangle$
is maximized again for $x^2=0$ and $z^2 = 1 - 2 s$, and it coincides with the ideal case:
$\langle d\ln s \rangle_{\parallel} = - 2\Gamma_0 dt\, \{ 2 - 2 s \}$, see Eq.~(\ref{time_pure1}).
On the other hand the change due to measurement in the complementary directions ($z=0$)
is deterministic and read:
$\langle d\ln s \rangle_{\perp} = d\ln s_{\perp} = - 2\Gamma_0 dt\,   [\frac{1}{\eta} + (1-\frac{1}{\eta})/{2s} ]$,
so that $\langle d\ln s \rangle\mid_{\rm   iso} = \langle d\ln s \rangle_{\parallel} + 2\,\langle d\ln s \rangle_{\perp}$.
By considering highly ideal detectors, $\eta \approx 1$, so that the last, singular, term in $d\ln s_{\perp}$ can be neglected,
one can reproduce the mean time $\langle T\rangle$ by integration of the equation for $\langle d\ln s \rangle\mid_{\rm   iso}$.
However, in the high-purity limit this is not possible: naive integration of log-entropy,
with the singular term, $\propto (1-\frac{1}{\eta})/{  s}$, included,
will lead to a wrong result.
Since in the high-purity limit $\delta \ll 1$, $\varepsilon  \ll 1$,
this term is of the order of $\frac{\delta}{\varepsilon}$,
%
it is clear that the time of reaching a certain purity level will be essentially
affected  when $\frac{\delta}{\varepsilon} \gtrsim 1$.

In what follows, we investigate      
the
   exact
solution for the MTFP in case of inefficient detectors.
We consider an initially completely mixed state, and denote the MTFP as
$\bar{T}(\delta,\varepsilon)$.
The FPE constructed above, based on the stochastic   It\^{o}   Eq.~(\ref{Ito-dp-3det}) is used,
where in the high-purity limit,
we approximate the coefficient,   Eq.~(\ref{FP-B-coeff}),   as:
$B(p)\simeq B_{\rm   HP}(p) = 8\Gamma_0\, (1-p)^2 $, while keeping $A(p)$ unchanged.
The MTFP solution  for one reflective boundary (at $p=1/2$) and one absorptive (at $p=1-\varepsilon$)
is written as (see, e.g. Ref.~\cite{Gardiner})
\begin{equation}
{  \bar{T}}(\delta,\varepsilon) = 2   \int_{p_0}^{1-\varepsilon}   \frac{dy}{\psi_{\rm   HP}(y,1-\delta)}
\int_{1/2}^{y} dz \frac{\psi_{\rm   HP}(z,1-\delta)}{B_{\rm   HP}(z)}
\label{MTFP} ,
\end{equation}
  where  $p_0$ is the initial purity as before;
in what follows we consider only $  p_0  =1/2$.
The function that enters the solution is
$\psi_{\rm   HP}(x)=\exp{\int_{1/2}^x\,  dx' \frac{2A(x')}{B_{\rm   HP}(x')}}$ and in our case
one obtains
\begin{eqnarray}
&& \psi_{\rm   HP}(x,1-\delta) = \frac{\exp{(x-\frac{1}{2})\, f_{1-\delta}(x)}}{[2(1-x)]^{\frac{1}{(1-\delta)}}}
\label{psi-MTFP}  ,
\end{eqnarray}
where we denoted $f_{1-\delta}(x) \equiv 1 - \frac{\delta}{(1-x) (1-\delta)}$.

For ideal measurements, $\delta=0$, one can integrate Eq.~(\ref{MTFP}) to obtain    analytically the result:
\begin{equation}
{  \bar{T}}(0,\varepsilon) \approx \frac{1}{8\Gamma_0} \left( \ln{\varepsilon^{-1}} - 1.35 \right)
\label{MTFP-delta0}  ,
\end{equation}
in agreement \cite{actual-B}
%
%
%
with the straightforward integration of the ensemble-averaged log-entropy equation,
[see Eqs.(\ref{Ito-dlog_s3})   and   (\ref{mean-time-3det}) above].
For $\delta > 0$ analytical integration is not as obvious: naive expansion of the integrand
in series in $\delta$ leads to a sum of contributions with increasing singularities.
We present instead numerical calculations that show the scaling behavior in the high-purity limit.

The main observation is that in the high-purity limit the time of purification can be
represented as
\begin{equation}
{  \bar{T}}(\delta,\varepsilon) \simeq
{  \bar{T}}(0,\varepsilon) + \Delta{  \bar{T}}\left(\frac{\delta}{\varepsilon}\right)
\label{scaling} .
\end{equation}
  That is,   the time increase $\Delta{  \bar{T}}$ depends only on the   ratio   $a \equiv \frac{\delta}{\varepsilon}$.
It can be shown that for   $a \ll 1$,
$\Delta{  \bar{T}}\left( a \right) \simeq  (1/8\Gamma_0)  \frac{a}{6}$,   while for   $a \gg 1$,
$\Delta{  \bar{T}}\left( a \right) \approx  (1/8\Gamma_0)  \exp{[ C_1(a)\, a ]}$,
with    $0.25 \lesssim C_1(a) < 0.5$,   approaching 0.5 for large   $a$.

\begin{figure}
\vspace*{-0.3cm} \centering
\includegraphics[width=3.7in]{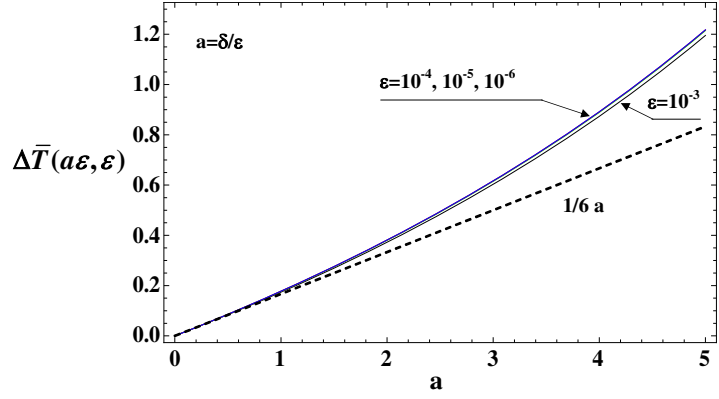}
\vspace*{-0.2cm}
\caption{The mean time increase, $\Delta\bar{T}(a\varepsilon,\varepsilon)$, for small $a < 5$, and $\varepsilon \ll 1$;
Scaling is established for $\varepsilon \lesssim 10^{-4}$.}
\label{Fig3}
\end{figure}

\begin{figure}
\vspace*{-0.3cm} \centering
\includegraphics[width=3.9in]{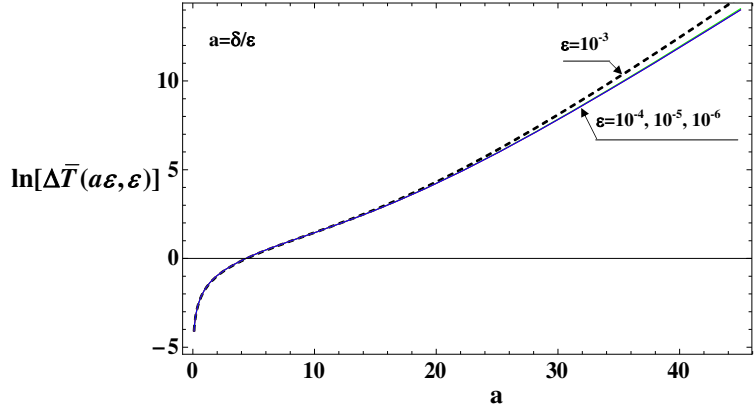}
\vspace*{-0.2cm}
\caption{The approximate linear growth of the logarithm of mean time increase,
$\ln[\Delta\bar{T}(a\varepsilon,\varepsilon)] \approx C_1 a$,
for large $a \gg 1$, and $\varepsilon \ll 1$;
Scaling is for $\varepsilon \lesssim 10^{-4}$.}
\label{Fig4}
\end{figure}

 On  Figs.~\ref{Fig3} and \ref{Fig4}  we present   evidence of scaling.
 Fig.~3 shows
$\Delta\bar{T}(a) = {  \bar{T}}(a \varepsilon,\varepsilon) - {  \bar{T}}(0,\varepsilon)$
at fixed $\varepsilon$ and for  $a < 5$.
 The curve for $\varepsilon = 10^{-3}$ is close to the
curves for $\varepsilon = 10^{-4},10^{-5},10^{-6}$ that practically coincide on the figure;
the slope at $a=0$ is $1/6$.
  Fig.~4   shows the scaling for large $a$.
 The $\ln[\Delta\bar{T}(a)]$,  eventually approaches a single curve
for $\varepsilon \ll 1$. For $a \gtrsim 20$ it looks like a linear growth, i.e.
$\ln[\Delta\bar{T}(a)] \approx C_1 a$.
More precisely, by numerical calculation of the derivative $d\ln[\Delta\bar{T}(a)]/da $ (not shown)
one can see that it
  is not a constant but varies somewhat as discussed above.
For small enough $\varepsilon$ ($a > 2000$),  $C_1$ approaches $1/2$.
Thus,   the   numerical calculations support Eq.~(\ref{scaling}).

\section{Conclusion}

In this paper we investigate  in detail a   recently proposed   quantum state purification protocol
based on simultaneous continuous measurement of the  three complementary observables
for a qubit: $\hat{\sigma}_x$, $\hat{\sigma}_y$, $\hat{\sigma}_z$.
Contrary to analogous single-detector purification protocols,
that are based on   complementarity \cite{Jacobs,WisemanRalph,ComJac06,ComWis11},
and which require feedback control,   here there is no need of introducing quantum feedback.
However, interestingly enough, the purification   dynamics   of our protocol can be
understood via the above mentioned protocols, using the established
isotropy of the qubit evolution under three complementary measurements by
identical detectors   \cite{RuskovKorotkovMolmer}.
%

For ideal   (i.e.   quantum-limited, or efficient) measurements, our main results are the observation of
  different factors for the speed-up (relative to a single measurement in the eigenbasis of the qubit state),
of 4 and 2, depending on how the purification speed is quantified.
In the first case it is  in terms of the  time
when the average system purity, $\langle p \rangle$, reaches certain purity level $1-\varepsilon$;
in the second case it is in terms of  the   mean time $\langle T \rangle$ at which the purity $p(T)$ first attains
the set level  of $1-\varepsilon$. Both of these speed-ups can be   understood via the possibility to
``split'' the three detector measurement at each measurement time step into   three equivalent
measurements --- one parallel to the state (in the Bloch sphere sense) and the remaining two perpendicular
to the state --- which is
one of the consequences of isotropy of the qubit evolution in the Bloch space.
  For the first measure of purification time these correspond
    to
  one ``bad'' and two ``good'' measurements
(giving speed-up contributions of $0$ and $2+2$ respectively, totalling 4). For the second measure,
they correspond to one ``good'' and two ''bad'' measurements (giving
 speed-up contributions of  $1$ and $1/2+1/2$ respectively, totalling 2).

For a measurement with non-ideal  detectors, the dynamics remains isotropic in the Bloch space. Moreover,
the classification to ``good'' and ``bad'' measurements
remains   the same, as long as the detector inefficiency $\delta$ is not greater than $1/2$.
The inefficiency causes an increase in the   purification times, that scales as a function of the ratio of
inefficiency over impurity, $\delta/\varepsilon$, in the high-purity and
high-efficiency   limit, $\delta \ll 1$, $\varepsilon \ll 1$.
Here, the first speed-up quantification, stated in terms of attaining a certain average purity,
is simply impossible if  $\varepsilon < \frac{\delta}{2}$,    
and the time required diverges as $\varepsilon$ approaches this limit  from above. On the other hand, the second quantification,
stated in terms of the mean time for individual (stochastically evolving) systems to attain this purity, always
gives an answer. However   we show that the mean time increases exponentially: $\propto \exp{[ 0.5\, \delta/\varepsilon]}$
for very large $\delta/\varepsilon$ ratio.
Still, for moderate $\delta/\varepsilon \gtrsim 1$ the mean time is   not too much greater than
the ideal case estimation.

\acknowledgments
The authors thank Alexander N. Korotkov  for useful discussions.
{RR was supported by Lundbeck foundation. KM was supported by the EU integrated project AQUTE. }
{ HMW and JC   were  supported by the Australian Research Council Centre of Excellence   CE110001027.
  JC also acknowledges support from National Science Foundation Grant No. PHY-0903953 \& PHY-1005540,
as well as Office of Naval Research Grant No. N00014-11-1-008.}


\begin{thebibliography}{99}
\bibitem{pre-existing}
An important difference is that a classical Bayesian inference
reveals a pre-existing system  state (e.g. a ``true'' probability distribution),
while quantum evolution, Eqs. (1), (2), is about the (final) quantum state available to an experimentalist.
\bibitem{actual-B}
Using the actual FPE coefficient $B(p)$,   Eq.~(\ref{FP-B-coeff}), numerical integration of
Eq.~(\ref{MTFP}) gives for the MTFP, $\bar{T}(0,\varepsilon) \approx \frac{1}{8\Gamma_0} \left( \ln{\varepsilon^{-1}} - 0.307 \right)$,
which differs negligibly from Eq.(\ref{mean-time-3det}) for large $\varepsilon^{-1}$.
\bibitem{note1}
In the absence of the two ``good''  measurements  the noise term must be taken into account to calculate
 the average purity $\langle p\rangle$.
\end{thebibliography}
\end{document}